\documentclass[a4]{paper}
\usepackage[utf8]{inputenc}
\usepackage[T2A]{fontenc}
\usepackage[english]{babel}
\usepackage{amssymb}
\usepackage{amsmath}
\usepackage{textcomp}
\usepackage{indentfirst}
\usepackage{float}
\usepackage{caption}
\usepackage{textcomp}
\DeclareCaptionLabelSeparator{pipe}{ $: $ }% or $\vert$
\usepackage{authblk}

\usepackage{units}
\usepackage{amsfonts}
\usepackage{graphicx} % ���������� ��������
\usepackage{amstext} % ��� ����������� �������� � �������������� ��������
\usepackage{wasysym} % ��� ����������

\usepackage{comment}
\usepackage{color}

\usepackage{epstopdf}
\usepackage{xcolor}
\usepackage{hyperref}
 
\definecolor{linkcolor}{HTML}{000000} % ���� ������
\definecolor{urlcolor}{HTML}{000000} % ���� �����������
 
\hypersetup{pdfstartview=FitH,  citecolor=black, linkcolor=linkcolor,urlcolor=urlcolor, colorlinks=true}
\usepackage[left=2.5cm,right=2cm, top=2cm,bottom=2cm,bindingoffset=0cm]{geometry} % ����

\DeclareMathOperator{\Sign}{sign}

\DeclareMathOperator{\arctanh}{arctanh}

\DeclareMathOperator{\Det}{Det}

\usepackage[sorting=none,backend=bibtex,citestyle=numeric]{biblatex}
\bibliography{literatureShock.bib}
 
\begin{document}

\title{Shock Waves in Relativistic Anisotropic Hydrodynamics}

\author{Aleksandr Kovalenko}
\author{Andrey Leonidov}
\affil{P.N. Lebedev Physical Institute, Moscow, Russia}

\maketitle

\begin{abstract}
Shock wave solutions in anisotropic relativistic hydrodynamics are analysed. A new phenomenon of anisotropy-related angular deflection of the incident flow by the shock wave front is described. Patterns of velocity and momentum transformation by the shock wave front are described. 
\end{abstract}

\section{Introduction}

The physics of ultrarelativistic heavy ion collisions is to a large extent determined by that of the Little Bang - evolution of hot and dense predominantly gluon matter created at the initial stage of these collisions, see \cite{Gelis:2021zmx} for a recent review. One of the characteristic features of this early evolution is a large pressure anisotropy due to formation of glasma flux tubes  \cite{Lappi:2006fp}. A standard way of describing expansion, cooling and subsequent transformation into final hadrons is to use the framework of relativistic dissipative hydrodynamics, see e.g. \cite{Baier:2006um,Romatschke:2009im,Jeon:2016uym}. The large difference between longitudinal and transverse pressure leads to a necessity of going beyond viscous hydrodynamics by summing over velocity gradients to all orders. A candidate theory of this sort is relativistic anisotropic hydrodynamics, see e.g. \cite{MartStr, RybFlor} and the review papers  \cite{Strickland,Alqahtani:2017mhy,Alqahtani:2018fcz} covering its both theoretical and phenomenological aspects.  A non-additive generalisation of relativistic anisotorpic hydrodynamics was recently suggested in \cite{Leonidov:2021jzi}. To analyse possible physical consequences of the pressure anisotropy it is of interest to study the  anisotropic versions of specific phenomena such as sound propagation  and shock waves. 

Sound propagation and Mach cone formation in anisotropic relativistic hydrodynamics was considered in \cite{Kirakosyan:2018afm}.  The present paper is devoted to the analysis of shock wave solutions in  anisotropic relativistic hydrodynamics. The shock wave solutions in relativistic hydrodynamics are known for a long  time, see e.g. \cite{landau2013course,Israel,Mitchell}. In applications to heavy ion physics the effects of shock waves were mostly discussed for low energy collisions  \cite{Scheid:1974zz,Gleeson:1979cb,Gleeson:1981nn}. An important exception is the study of \cite{Gyulassy:1996ka,Gyulassy:1996br} of transverse shock waves generated in the primordial turbulent gluon/minijet medium in high energy heavy ion collisions. With modern glasma type understanding of the essentially anisotropic nature of this medium it is of interest to rethink the results of \cite{Gyulassy:1996ka,Gyulassy:1996br} in terms of transverse shocks generated in anisotropic relativistic hydrodynamics. The present paper is a first step in this direction.

\section{Shock waves in anisotropic relativistic hydrodynamics}

\subsection{Shock wave discontinuity in isotropic relativistic hydrodynamics}

In this section we set the framework of subsequent analysis by reminding of the necessary information on shock wave discontinuity in isotropic relativistic hydrodynamics \cite{landau2013course,Israel,Mitchell}. We focus on the shock wave solution in the ideal fluid characterised by the energy-momentum tensor
\begin{equation}
T_{\mu\nu} = (\varepsilon + P) U^\mu U^\nu - P g^{\mu nu}
\label{tmnid}
 \end{equation}
where $\varepsilon$ is energy density, $P$ - pressure and $U^\mu$ is the four-vector of the flow velocity satisfying $U_\mu U^\mu = 1$. In this case, the shock wave is described by a discontinuous solution of the equations of motion such that components of energy-momentum tensor normal to the discontinuity hypersurface are discontinuous across it while tangential ones remain continuous. 

The energy-momentum conservation then leads to the following matching condition linking downstream and upstream projections on the direction perpendicular to the discontinuity surface:
\begin{equation}
T_{\mu\nu} N^\mu = T^{'}_{\mu\nu} N^\mu,
\label{gap}
\end{equation}
where  \(N^\mu\) - unit  vector normal to the discontinuity surface and  \( T_{\mu\nu} \) and \( T^{'}_{\mu\nu} \) correspond to upstream and downstream energy-momentum tensors correspondingly.  

A quantitative description of a shock wave is that of a transformation of pressure, entropy $S$ and normal component of velocity $v$ across the shock wave surface:
\begin{equation}
(P,S,v) \;\; \Rightarrow \;\; (P',S',v')
\end{equation}
In this paper we will consider only the case of a compression shock wave for which $P'>P$, $S'>S$ and $v'<v$ (see a detailed derivation of the expression for the velocity drop below). The ratio $\sigma=P'/P$ will be considered as a parameter characterising the shock wave solution.

Using the explicit expression the energy-momentum tensor \eqref{tmnid}  we get
\begin{equation}
(\varepsilon + P)U_\mu N^\mu U_\nu - P N_\nu = (\varepsilon^{'} + P^{'})U^{'}_\mu N^\mu U^{'}_\nu  - P^{'} N_\nu.
\label{gap_open}
\end{equation}
Taking the product of equation (\ref{gap_open}) with \(U^\nu\) and \(U^{'\nu}\) one can obtain the following system:
\begin{align}
(\varepsilon + P^{'}) x = (\varepsilon^{'} + P^{'}) A x^{'},
\label{n1}
\\
(\varepsilon^{'} + P) x^{'} = (\varepsilon + P) A x ,
\label{n2}
\end{align}
where we have defined  \(x=U_\mu N^\mu \), \( x^{'} = U^{'}_\mu N^\mu \) and \(A = U^{'}_\nu U^\nu\). We get
\begin{align}
(\varepsilon + P^{'}) (\varepsilon + P) x^2 = (\varepsilon^{'} + P) (\varepsilon^{'} + P^{'}) x^{'2}.
\label{xx}
\end{align}
The vector \(N^\mu\) must be space-like, \(N^\mu N_\mu <0 \), for discontinuity surface to propagate inside the light cone and thus be subluminal  \cite{Israel}. After multiplying equation (\ref{gap_open}) by \(N_\mu\) one finds
\begin{equation}
N_\nu N^\nu = \frac{1}{P - P^{'}} \Big[ (\varepsilon + P)x^2 - (\varepsilon^{'} + P^{'}) x^{'2} \Big],
\label{N_norm}
\end{equation}
and, using (\ref{xx}) one gets
\begin{align}
N_\mu N^\mu &= \frac{\varepsilon^{'} + P^{'}}{\varepsilon + P^{'}} \Big[ 1 - \frac{\varepsilon - \varepsilon^{'}}{P - P^{'}} \Big] = \frac{\varepsilon + P}{\varepsilon^{'} + P} \Big[ 1 - \frac{\varepsilon - \varepsilon^{'}}{P - P^{'}} \Big].
\label{N_norm2}
\end{align}
The subliminality condition \(N_\mu N^\mu<0\) is thus insured by the following inequality
\begin{align}
\frac{\varepsilon - \varepsilon^{'}}{P - P^{'}} > 1.
\label{inequality}
\end{align}
In ultra-relativistic case, then \(\varepsilon = 3P\), the inequality \eqref{inequality} is trivially satisfied.

For the discussion below it is useful to remind an expression for the upstream and downstream velocities \cite{Mitchell} in the ultrarelativistic case.  Choosing \(N^\mu = (0, 1, 0, 0)\) and 4-velocity vector of the form \(U^\mu = (U^0, U_x, 0, 0)\) we get:
\begin{align}
4 P U_x U_0 & = 4 P^{'} U^{'}_x U^{'}_0,
\label{iso_system_1}
\\
- 4 P U_x^2 - P & = - 4 P^{'} U^{'2}_x  - P^{'}.
\label{iso_system_2}
\end{align}
so that in terms in terms of the velocity components \(v_i = U_i/U_0\) 
\begin{align}
v_x = \sqrt{\frac{3\sigma + 1}{3(\sigma + 3)}}, \, \, v_x^{'} = \sqrt{\frac{\sigma + 3}{3(3\sigma + 1)}}.
\label{iso_solution_x}
\end{align}
 %An important characteristic of a shock wave is a relative difference between upstream and downstream velocities
%\begin{equation}
%\label{defdelta}
%\delta = \frac{v-v'}{v}
%\end{equation}
A compact characterisation of the velocity transformation \eqref{iso_solution_x} across the shock wave front is given by relative difference between upstream and downstream velocities 
\begin{equation}
\label{deltahom}
\delta_{\rm iso}  = \frac{v'_x-v_x}{v_x} =-\frac{2}{3 \sigma+1}  (\sigma-1) 
\end{equation}
For the considered case of a compression shock wave $P'>P$ one has $\sigma>1$ and, therefore, it follows from \eqref{deltahom} that $\delta_{\rm iso} <0$ so that the flow velocity indeed drops across the compression shock wave front.   
Let us also note that 
\begin{equation}
\label{velprod}
\rho_{\rm iso} = v_x v'_x = \frac{1}{3}=c^2_s
\end{equation}
where $c_s$ is a speed of sound.

One of the main topics of the analysis below is the one of the velocity transformation ${\bf v} \to {\bf v'}$ in anisotropic relativistic hydrodynamics generalising the formulae \eqref{deltahom} and \eqref{velprod} for the isotropic case. 

\subsection{Anisotropic relativistic hydrodynamics}

Our treatment of anisotropic relativistic hydrodynamics will follow the kinetic theory - founded approach of  \cite{MartStr,StrRom1,StrRom2} based on working with a specific ansatz for a distribution function
\begin{equation}
f(x,p) = f_{iso}\Bigg( \frac{\sqrt{p^\mu \Xi_{\mu\nu}(x) p^\nu}}{\Lambda(x)}\Bigg),
\label{Kin1}
\end{equation} 
where $\Lambda(x)$ is a coordinate-dependent temperature-like momentum scale and $\Xi_{\mu\nu}(x)$ quantifies  coordinate-dependent momentum anisotropy. In what follows we consider one-dimensional anisotropy so that   $ (p^\mu \Xi_{\mu\nu}p^\nu = \mathbf{p}^2 +\xi(x) p_\parallel^2)$  in the local rest frame (LRF). To obtain a transparent parametrisation of the energy-momentum tensor in anisotropic hydrodynamics it is convenient to rewrite the four-vector $U^\mu (x)$ in terms of the longitudinal rapidity $\vartheta (x)$, the timelike velocity $u_0$ and transverse velocities $u_x, u_y$
\begin{equation}
U^\mu = (u_0 \cosh \vartheta, u_x, u_y, u_0 \sinh \vartheta)
\end{equation}
where $u_0^2=1+u_x^2+u_y^2$ and define a space-like unit vector
\begin{equation}
Z^\mu = (\sinh \vartheta, 0,0, \cosh \vartheta)
\end{equation}
such that $Z^\mu Z_\mu = -1$ which is orthogonal to $U^\mu$, $Z_\mu U^\mu=0$. 
  
Using a standard definition for energy-momentum tensor as the second moment of the distribution function
\begin{equation}
T^{\mu\nu} = \int \frac{d^3 p}{(2\pi)^2 p_0} p^\mu p^\nu f_{iso}\Bigg(\frac{\sqrt{p^\mu \Xi_{\mu\nu}(x)p^\nu}}{\Lambda(x)}\Bigg)
\label{T}
\end{equation}
one can derive the following equation for the energy-momentum tensor $T^{\mu\nu}$:
\begin{equation}
T^{\mu\nu} = (\varepsilon + P_\perp)U^\mu U^\nu - P_\perp g^{\mu\nu} + (P_\parallel - P_\perp)Z^\mu Z^\nu,
\label{T_true}
\end{equation}
$P_\parallel$ and  $P_\perp$ is longitudinal (towards anisotropy direction) and transverse pressure respectively. In the LRF the expression \eqref{T_true} takes the form
\begin{equation}
T^{\mu\nu} = {\rm diag} (\varepsilon,P_\perp,P_\perp,P_\parallel)
\label{Tlrf}
\end{equation}
Let us note that in the ultra-relativistic case the condition of the tracelessness of the energy-momentum tensor leads to the relation $\varepsilon = 2P_\perp + P_\parallel$.

The dependence on the anisotropy parameter $\xi$ can be factorised so that
\begin{align}
\varepsilon &= \int \frac{d^3 p}{(2\pi)^2} p^0 f_{iso}\Bigg(\frac{\sqrt{\textbf{p}^2 + \xi(x) p_\parallel^2}}{\Lambda(x)}\Bigg) = R(\xi) \varepsilon_{iso} (\Lambda),
\label{e} \\
P_\perp &= \int \frac{d^3 p}{(2\pi)^2} \frac{p_\perp^2}{2 p_0}f_{iso}\Bigg(\frac{\sqrt{\textbf{p}^2 + \xi(x) p_\parallel^2}}{\Lambda(x)}\Bigg) = R_\perp (\xi) P_{iso} (\Lambda),
\label{pT} \\
P_\parallel &= \int \frac{d^3 p}{(2\pi)^2} \frac{p_\parallel^2}{p_0}f_{iso}\Bigg(\frac{\sqrt{\textbf{p}^2 + \xi(x) p_\parallel^2}}{\Lambda(x)}\Bigg) = R_\parallel (\xi) P_{iso} (\Lambda),
\label{pL}
\end{align}
where the anisotropy-dependent factors $R_\perp(\xi)$ and $R_\parallel(\xi)$ read \cite{StrRom1,StrRom2}
\begin{equation}
\label{RTL}   
R_\perp(\xi) = \frac{3}{2\xi} \Bigg( \frac{1 + (\xi^2-1)R(\xi)}{1+\xi}\Bigg), \;\;\;\;\;\;
R_\parallel(\xi) = \frac{3}{\xi} \Bigg( \frac{(\xi+1)R(\xi) -1}{1+\xi}\Bigg), 
\end{equation}
where, in turn, 
\begin{equation}
\label{R}
  R(\xi) = \frac{1}{2} \Bigg( \frac{1}{1+\xi} + \frac{\arctan \sqrt{\xi}}{\sqrt{\xi}}\Bigg). 
\end{equation}
Let us note that the anisotropy factors (\ref{RTL},\ref{R}) are related by the following useful formula:
\begin{equation}
\label{RT+RP}
2 R_\perp(\xi) + R_\parallel(\xi) = 3 R(\xi)
\end{equation}

In the preceding paper \cite{Kirakosyan:2018afm} we have derived the following equation describing propagation of sound in relativistic anisotropic hydrodynamics with longitudinal anisotropy:
\begin{equation}
\partial^2_t \;  n^{(1)} = \left ( c^2_{s \perp} \; \partial^2_\perp + c^2_{s \parallel} \; \partial^2_z \right)  n^{(1)} 
\end{equation}
where $n^{(1)}$ is a (small) density fluctuation and $c_{s \perp}$ and $c_{s  \parallel}$ stand for anisotropy-dependent transverse and longitudinal speed of sound respectively. The explicit expressions for $c^2_{s \perp}$ and $c^2_{s \parallel}$ read  \cite{Kirakosyan:2018afm}:
\begin{equation}
\label{csTR1}
 c^2_{s \perp} =  \frac{R_\perp}{2 R_\perp + R_\parallel}, \;\;\; c^2_{s \parallel} = \frac{R_\parallel}{2 R_\perp + R_\parallel}. 
\end{equation}
Using equation \eqref{RT+RP} the expressions \eqref{csTR1} can be rewritten in the following simple form:
\begin{equation}
\label{csTR2}
c^2_{s \perp}  =  \frac{R_\perp}{3 R}, \;\;\; c^2_{s \parallel} = \frac{R_\parallel}{3 R}
\end{equation}

\subsection{Transverse and longitudinal shock waves}

\subsubsection{Transverse normal shock wave}\label{tsw}

Let us first consider a description of a transverse shock wave. Due to the symmetry in \(Oxy\)-plane, it sufficient to consider its propagation along the $x$ axis and, correspondingly, choose the following basis:
\begin{align}
U_\mu &= (u_0, u_x, 0, 0), \, \, \, Z_\mu = (0, 0, 0, 1),
\label{basis1}
\\
U^{'}_\mu &= (u^{'}_0, u^{'}_x, 0, 0), \, \, \, Z^{'}_\mu =  Z_\mu = (0, 0, 0, 1).
\label{basis2}
\end{align}
Similarly to the example from relativistic hydrodynamics, consider the case when the normal vector is directed along the \(Ox\)-axis \(N^\mu = (0,1,0,0 )\) (solution for the case of an arbitrary \(N^\mu\) see Appendix B). 
In the ultrarelativistic case The matching conditions (\ref{gap}) then lead to the following system of equations: 
\begin{align}
& (3P_\perp + P_\parallel)u_0  u_x - (3P^{'}_\perp + P^{'}_\parallel) u^{'}_0\, u^{'}_x  = 0,
\label{eqs2_1_T}
\\
& (3P_\perp + P_\parallel)u_x^2    + P_\perp  -(3P^{'}_\perp + P^{'}_\parallel) (u^{'}_x)^2  - P^{'}_\perp = 0.
\label{eqs2_2_T}
\end{align}

From  the equations (\ref{eqs2_1_T} - \ref{eqs2_2_T}) we find expressions for the velocities \(u_x, \ u_x^{'}\):
\begin{align}
& u_x = \frac{v_x}{u_0} =  \sqrt{\frac{(P_\perp - P^{'}_\perp) (P_\parallel^{'} + P_\perp + 2P_\perp^{'})}{(2P_\perp - P_\parallel - 2P^{'}_\perp + P^{'}_\parallel)(P_\parallel^{'} + 2P_\perp + P_\perp^{'})}},
\label{u1}
\\
& u_x^{'} = \frac{v'_x}{u'_0} = \sqrt{\frac{(P_\perp - P^{'}_\perp)(P_\parallel^{'} + 2P_\perp + P_\perp^{'})}{(2P_\perp - P_\parallel - 2P^{'}_\perp + P^{'}_\parallel)(P_\parallel^{'} + P_\perp + 2P_\perp^{'})}},
\label{u2}
\end{align}

It is assumed that the anisotropy parameter does not change near the shock wave, since anisotropy is related to the properties of the medium, thus \(\xi^{'} = \xi\). Using the formulae (\ref{pT})-(\ref{R}) for the transverse and longitudinal pressure and anisotropy factors one gets the following expressions for the upstream and downstream velocities $v_x$ and $v_x^{'}$:
\begin{equation}
\label{velT}
v_x (\sigma, \xi) = \sqrt{\frac{R_\perp(3 \sigma R + R_\perp)}{3R(R_\perp\sigma + 3 R))}}, \;\;\; 
v_x^{'} (\sigma, \xi) = \sqrt{\frac{R_\perp (R_\perp\sigma + 3 R)}{3R(3 \sigma R + R_\perp)}},
\end{equation}
where, as before, \(\sigma = P^{'}_{iso} / P_{iso}\).
 
Using the expressions \eqref{velT} we can calculate the relative difference $\delta_\perp$ and the product $\rho_\perp$ of the upstream and downstream velocities
\begin{eqnarray}
\delta_\perp (\xi) & = & \frac{v'_x - v_x}{v_x} = -(\sigma - 1) \frac{3R - R_\perp}{3\sigma R + R_\perp} \label{delT} \\
\rho_\perp (\xi) & = & v_x v_x^{'}  =  \frac{R_\perp}{3R}  \label{rhoT}
\end{eqnarray}

Recalling the fact that the product of upstream and downstream velocities in the isotropic case is equal to the speed of sound squared, see equation \eqref{velprod}, one can identify such a product for the transverse shock wave with a transverse speed of sound squared
\begin{equation}
\label{ssT}
\rho_\perp (\xi) = c^2_{s \perp}
\end{equation}
Comparing equation \eqref{rhoT} and the first equation in \eqref{csTR2} we see that this definition leads to an expression identical to that following from the equation for sound propagation in relativistic anisotropic hydrodynamics derived in \cite{Kirakosyan:2018afm}.

In the isotropic limit $\xi \to 0$ 
\begin{equation}
\left. \delta_\perp (\xi) \right \vert_{\xi \to 0} \to \delta_{\rm iso}, \;\;\; \left. \rho_\perp (\xi) \right \vert_{\xi \to 0} \to \rho_{\rm iso},
\end{equation}
where $\delta_{\rm iso}$ and $ \rho_{\rm iso}$ were defined in equations \eqref{deltahom} and \eqref{velprod}.

In the opposite limit of $\xi \to \infty$
 \begin{equation}
 \label{drt}
\left. \delta_\perp (\xi) \right \vert_{\xi \to \infty} \to  -(\sigma-1)\frac{1}{2 \sigma+1} , \;\;\; \left. \rho_\perp (\xi) \right \vert_{\xi \to \infty} \to \frac{1}{2},
\end{equation}

The functions $\delta_\perp (\xi)$ and $\rho_\perp (\xi)$ are plotted in Figs.~(\ref{ris1},\ref{speed_sound}).

\subsubsection{Longitudinal normal shock wave}\label{lsw}

Similar calculations can be carried out for the longitudinal shock wave propagating along the anisotropy axis. In this case we choose
\begin{equation}
\label{basis}
U_\mu = (\cosh \vartheta, 0, 0, \sinh \vartheta),\;\;\;\;Z_\mu = (\sinh \vartheta, 0, 0, \cosh \vartheta).
\end{equation}
Proceeding analogously to the previously considered case for the transverse shock wave we get
\begin{equation}
v_z (\sigma, \xi) = \sqrt{\frac{R_\parallel(3\sigma R + R_\parallel)}{3R(R_\parallel\sigma + 3R))}}, \;\;\;
v_z^{'} (\sigma, \xi) = \sqrt{\frac{R_\parallel (R_\parallel\sigma + 3R)}{3R(3 \sigma R + R_\parallel)}}.
\end{equation}
The corresponding expressions for $\delta_\parallel (\xi)$ $\rho_\parallel (\xi)$ read
\begin{eqnarray}
\delta_\parallel (\xi) & = & \frac{v'_z - v_z}{v_z} = -(\sigma - 1) \frac{3R - R_\parallel}{3\sigma R + R_\parallel} \label{delL} \\
\rho_\parallel (\xi) & = & v_z v_z^{'} = \frac{R_\parallel}{3R} \label{rhoL}
\end{eqnarray}

Defining analogously to \eqref{ssT}
\begin{equation}
\label{ssL}
\rho_\parallel (\xi) = c^2_{s \parallel}
\end{equation}
and comparing equation \eqref{rhoL} with the second equation in \eqref{csTR2} we see that like in the transverse case this definition leads to an expression identical to that following from the equation for sound propagation in relativistic anisotropic hydrodynamics derived in \cite{Kirakosyan:2018afm}.

In the isotropic limit $\xi \to 0$ 
\begin{equation}
\left. \delta_\parallel (\xi) \right \vert_{\xi \to 0} \to \delta_{\rm iso}, \;\;\; \left. \rho_\parallel (\xi) \right \vert_{\xi \to 0} \to \rho_{\rm iso}.
\end{equation}

In the opposite limit of $\xi \to \infty$
 \begin{equation}
 \label{drl}
\left. \delta_\parallel (\xi) \right \vert_{\xi \to \infty} \to  -(\sigma-1)\frac{1}{ \sigma} , \;\;\; \left. \rho_\parallel (\xi) \right \vert_{\xi \to \infty} \to 0.
\end{equation}

The functions $\delta_\parallel (\xi)$ and $\rho_\parallel (\xi)$ are plotted, together with their counterparts $\delta_\perp (\xi)$ and $\rho_\perp (\xi)$, in Figs.~(\ref{ris1},\ref{speed_sound}).

\subsubsection{Comparison between normal transverse and longitudinal shock waves}\label{ctlsw}

\begin{figure}[H]
\center{\includegraphics[width=0.6\linewidth]{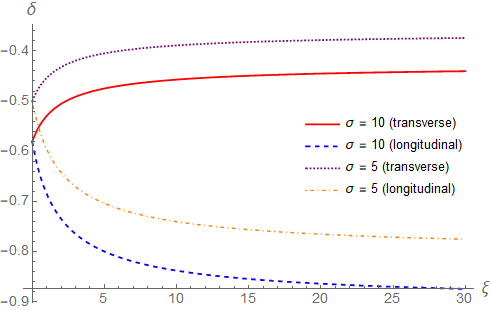}}
\caption{\small Plot of \(\delta(\xi \vert \sigma)\) for longitudinal ($\sigma=5$ - dash-dotted line, $\sigma=10$ - dashed line ) and transverse ($\sigma=5$ - dotted line , $\sigma=10$ - solid line) shock waves.}
\label{ris1}
\end{figure}

\begin{figure}[h]
\center{\includegraphics[width=0.64\linewidth]{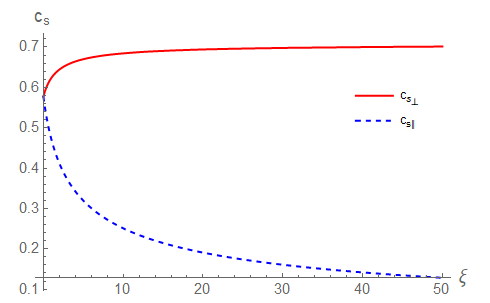}}
\caption{\small Plots of \(c_{s\perp}\) (solid) and \(c_{s\parallel}\) (dashed) in units of the speed of light.}
\label{speed_sound}
\end{figure}

From Figs.~(\ref{ris1},\ref{speed_sound}) we see that the anisotropy dependence of the relative rapidity drop  $\delta_\perp (\xi)$ and velocities product $\rho_\perp (\xi)$ and that of their longitudinal counterparts $\delta_\parallel (\xi)$ and $\rho_\parallel (\xi)$ are of different character. 

Starting from the same (negative) value $\delta_{\rm iso}$ at $\xi=0$, the relative velocity drop $\delta_\perp (\xi)$ grows with $\xi$ towards its asymptotic value given in \eqref{drt}. This means that  the  velocity gap for the transverse shock wave shrinks with growing $\xi$ so that the transverse shock wave weakens with increasing anisotropy. On the contrary, the relative velocity drop $\delta_\parallel (\xi)$ decays with $\xi$ towards it asymptotic value given in \eqref{drl} with, therefore, the relative velocity gap for the longitudinal shock wave widening with growing $\xi$ so that the longitudinal shock wave strengthens with increasing anisotropy.  At asymptotically large anisotropies $\xi \to \infty$ the gap between the transverse and longitudinal relative velocity drop reaches
\begin{equation}
\left.  \left( \delta_\perp (\xi) - \delta_\parallel (\xi) \right) \right \vert_{\xi \to \infty} \;\; \to \;\; \frac{(\sigma-1)^2}{\sigma(2 \sigma+1)}
\end{equation}

As to the anisotropy behaviour of the velocities product or, equivalently, the corresponding speed of sound, the transverse speed of sound  \(c_{s\perp}\) grows from $1/\sqrt{3}$ at \(\xi=0\) to \(1/\sqrt{2}\) at $\xi \to \infty$ while the longitudinal one  \(c_{s\parallel}\) decays from the same value $1/\sqrt{3}$ at \( \xi=0 \) to $0$ at $\xi \to \infty$. Since the existence of a shock wave is possible only when the flow moves with a velocity greater than the speed of sound, a much lower flow velocity is required for the shock wave generation in the direction of anisotropy. Therefore, for larger anisotropies formation of longitudinal shock waves is becomes progressively easier while that  of  transverse ones is, on the contrary, becoming more difficult.

\subsection{Normal shock wave at an arbitrary polar angle}

In this section we develop a description of a normal shock wave incident at an arbitrary polar angle. A major new element we are going to encounter is that, in contrast with the above-considered cases of transverse and longitudinal shock waves, the transformation ${\bf v} \to {\bf v'}$ of the upstream velocity to the downstream one in a shock wave incident at an arbitrary polar angle involves changes both in the absolute value of the flow velocity and in the direction of its propagation, see Fig.\;\ref{ris_scheme}.

To characterise the transformation ${\bf v} \to {\bf v'}$ we introduce the following variables describing changes in the absolute value and direction of flow velocity across the shock wave front:
\begin{equation}
\label{deltaalpha}
\delta_{\alpha \; \alpha'} (\xi) = \frac{\vert {\bf v'} (\xi \vert \alpha) \vert - \vert {\bf v} (\xi \vert \alpha') \vert}{\vert {\bf v (\xi \vert \alpha)} \vert }, \;\;\; \beta (\xi) = \alpha' (\xi) - \alpha
\end{equation} 
where $\delta_{\alpha \alpha'} (\xi) $ generalises the variables $\delta_\perp (\xi)$ and $\delta_\parallel (\xi)$ defined in \eqref{rhoT} and \eqref{rhoL} correspondingly.

\subsubsection{General equations}

\begin{figure}[H]
\center{\includegraphics[width=0.5\linewidth]{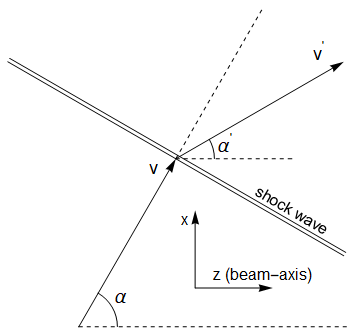}}
\caption{Transformation of flow velocity by the shock wave front}
\label{ris_scheme}
\end{figure}

For simplicity, but without loss of generality, consider a flow moving at an angle \(\alpha\) to the direction perpendicular to the beam axis (Figure \ref{ris_scheme}). Due to the symmetry in the \(Oxy\) plane one can choose, for instance, \(x\)-axis. The difference between angles \(\alpha^{'}\) and \(\alpha\) denote as \(\beta\).

Thus, the velocities expressed in terms of the 4-velocity vector components are
\begin{align}
v_x &= \frac{u_x}{u_0} = \frac{\tanh \gamma}{\cosh \vartheta} = v \cos \alpha, \;\;\;\;\;\; v_z = \frac{u_z}{u_0} = \tanh \vartheta = v \sin \alpha, \label{vxz1_ugol} \\
v_x^{'} &= \frac{u_x^{'}}{u_0^{'}} = \frac{\tanh \gamma^{'}}{\cosh \vartheta^{'}} = v^{'} \cos \alpha^{'}, \;\;\;
v_z^{'} = \frac{u_z^{'}}{u_0^{'}} = \tanh \vartheta^{'} = v^{'} \sin \alpha^{'} \label{vxz2_ugol}.
\end{align}
From equations (\ref{vxz1_ugol},\ref{vxz2_ugol}) one gets the following expressions for  \(\vartheta, \vartheta^{'} \):
\begin{equation}
\label{var12}
\vartheta = \arctanh \big[ \tanh \gamma \tan \alpha \big], \;\;\; \vartheta^{'} = \arctanh \big[ \tanh \gamma^{'} \tan \alpha^{'} \big].
\end{equation}

Let us choose the following parametrisation for the components of the  vector normal to the discontinuity surface:
\begin{align}
N_\mu = (0, \cos \alpha,  0, \sin \alpha), 
\label{n_ugol}
\end{align}

With the parametrisation \eqref{n_ugol} the matching conditions \eqref{gap} take the following form:
\begin{align}
& -(3P_\perp + P_\parallel)\cosh \gamma \cosh \vartheta \, ( \sinh \gamma \cos \alpha + \cosh \gamma \sinh \vartheta \sin \alpha)  + \nonumber \\ 
& + (3P^{'}_\perp + P^{'}_\parallel)\cosh \gamma^{'} \cosh \vartheta^{'} \,  ( \sinh \gamma^{'} \cos \alpha + \cosh \gamma^{'} \sinh \vartheta^{'} \sin \alpha)  + \nonumber \\ 
& + (P_\perp - P_\parallel)\sinh \vartheta \cosh \vartheta \sin \alpha- (P^{'}_\perp - P^{'}_\parallel)\sinh \vartheta^{'} \,  \cosh \vartheta^{'} \sin \alpha = 0, 
\label{eqs_ugol_1} \\
\nonumber \\
& - (3P_\perp + P_\parallel)\cosh \gamma \sinh \vartheta \,  (\sinh \gamma \cos \alpha + \cosh \gamma \sinh \vartheta \sin \alpha)  - P_\perp \sin \alpha +\nonumber \\ 
 & + (3P^{'}_\perp + P^{'}_\parallel)\cosh \gamma^{'} \sinh \vartheta^{'} \, (\sinh \gamma^{'} \cos \alpha + \cosh \gamma^{'} \sinh \vartheta^{'} \sin \alpha) + P^{'}_\perp \sin \alpha + \nonumber \\ 
& + (P_\perp - P_\parallel)\cosh \vartheta \, \cosh \vartheta \sin \alpha - (P^{'}_\perp - P^{'}_\parallel)\cosh \vartheta^{'} \, \cosh \vartheta^{'} \sin \alpha  = 0,
\label{eqs_ugol_2} \\
\nonumber \\
& - (3P_\perp + P_\parallel)\sinh \gamma  (\sinh \gamma \cos \alpha + \cosh \gamma \sinh \eta \sin \alpha)  - P_\perp \cos \alpha + \nonumber \\ 
& +(3P^{'}_\perp + P^{'}_\parallel)\sinh \gamma^{'} (\sinh \gamma^{'} \cos \alpha + \cosh \gamma^{'} \sinh \vartheta^{'} \sin \alpha) + P^{'}_\perp \cos \alpha = 0, 
\label{eqs_ugol_3}
\end{align}

The equations (\ref{eqs_ugol_1}) - (\ref{eqs_ugol_3}) constitute a system of equations for three unknowns   \(\gamma, \gamma^{'}, \alpha^{'}\)  that depend on 
three parameters \(\sigma, \xi, \alpha\) that is solved numerically. The formulae (\ref{vxz1_ugol},\ref{vxz2_ugol}) then translate a solution for   \(\gamma, \gamma^{'}, \alpha^{'}\) into vectors of upstream and downstream velocities.

\subsubsection{Flow deflection by the shock wave front}

In isotropic relativistic hydrodynamics, a normal shock wave changes only the absolute value of the incident flow velocity, but not the direction of the flow passing through it. Our analysis of transverse and longitudinal normal shock waves in paragraphs \ref{tsw} and \ref{lsw} has shown that in these cases deflection of the incident flow is also absent in anisotropic hydrodynamics. However, it turns out that this is no longer true for normal shock waves incident at an arbitrary polar angle.  The flow is deflected by the shock wave front so that in notations of Fig.~\ref{ris_scheme}  $\alpha^{'} \neq \alpha$.

\begin{figure}[H]
\center{\includegraphics[width=1\linewidth]{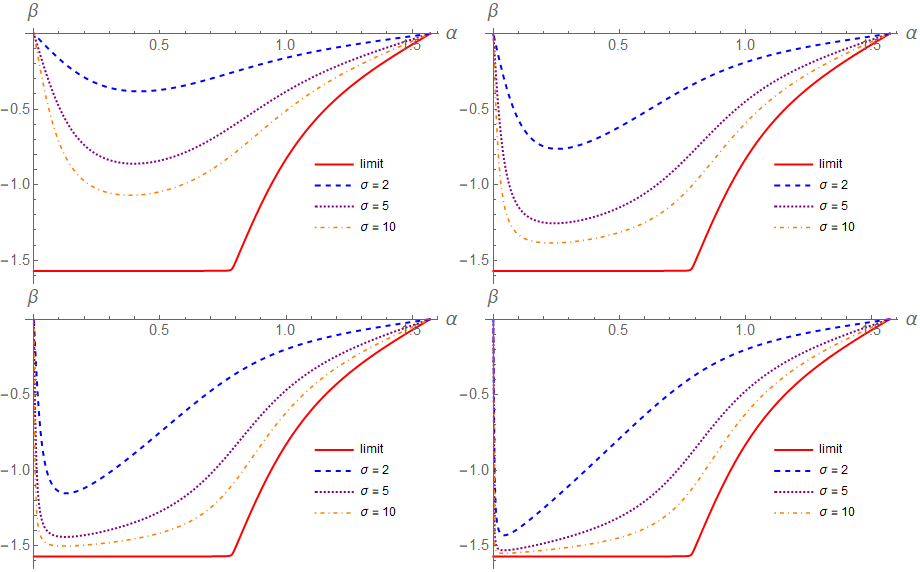}}
\caption{\small Plot of \(\beta = \alpha^{'} - \alpha\) as a function of \(\alpha\) for  $\sigma = 2$ (dashed line), $\sigma=5$ (dotted line) and $\sigma=10$ (dash-dotted line)  for  \(\xi = 5\) (top left), \(\xi = 20\) (top right), \(\xi = 100\) (bottom left), \(\xi = 1000\) (bottom right). Each plot contains the limiting curve \(\sigma \rightarrow \infty, \ \xi \rightarrow \infty\) (solid line).}
\label{ris_ugol}
\end{figure}

In Fig.~\ref{ris_ugol} we plot the deflection angle  \(\beta = \alpha^{'} - \alpha\) as a function of the incidence angle $\alpha$ for several values of $\sigma$ and different anisotropies. In all the cases the function $\beta (\alpha)$ takes negative values  and has a minimum at some $\alpha^*$. At fixed $\xi$ the depth of this minimum grows with $\sigma$. At fixed $\sigma$ with growing $\xi$ the minimum a) gets deeper and b) its position shifts towards smaller $\alpha$. Let us note that for strong anisotropy, large \(\sigma \) and incidence angles  \(\alpha \leq \pi / 4 \) we have a flow deviation from the initial direction by almost \(\pi/2\) so that the upstream flow tends to propagate along the shock wave front. 
Let us note that from the bottom plots of Fig.~\ref{ris_ugol} we see that for large anisotropies one observes a rapid change of regimes indicating an existence of effective instability at small angles.

\subsubsection{Transformation $\vert {\bf v (\xi)} \vert \to \vert {\bf v' (\xi)} \vert$}\label{vvp}

\begin{figure}[H]
\center{\includegraphics[width=1\linewidth]{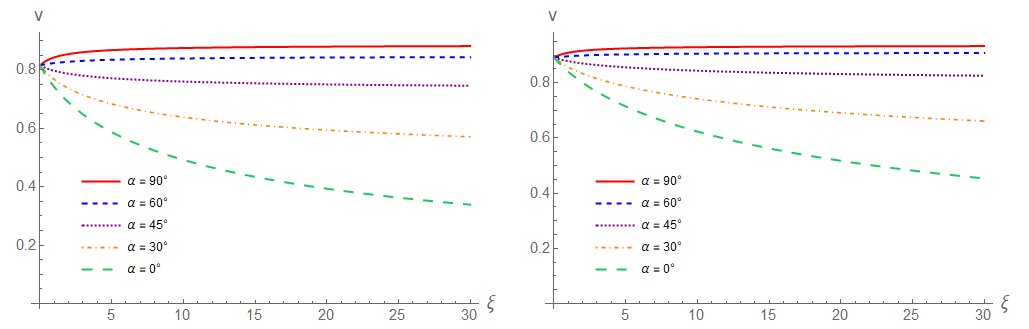}}
\caption{\small Plots of \( v_\alpha (\xi) \) as a function of the anisotropy parameter \( \xi \) for $\alpha=\pi/2$ (solid),  $\alpha=2\pi/3$ (dashed), $\alpha=\pi/4$ (dotted), $\alpha=\pi/6$ (dash-dotted)  and $\alpha=0$ (long dash) for  \(\sigma = 5\) (left) and \(\sigma = 10\) (right).}
\label{graph_v1}
\end{figure}

Let us first consider the anisotropy dependence of the absolute value of upstream velocity $v_\alpha (\xi) \equiv \vert {\bf v (\xi \vert \alpha)} \vert $ at different incidence angles $\alpha$, $\alpha \in [0,\pi/2]$. The resulting curves  are shown in Fig.~\ref{graph_v1}  for two different values of $\sigma$.
In Fig.~\ref{graph_v1} we see, for both values of $\sigma$, a transition from convex decaying \( v_\alpha (\xi) \) at small incidence angles \(\alpha\)  to concave growing one at large incidence angles. The transition takes place at \(\alpha_{\rm crit} \sim \pi/4\).

\begin{figure}[H]
\center{\includegraphics[width=1\linewidth]{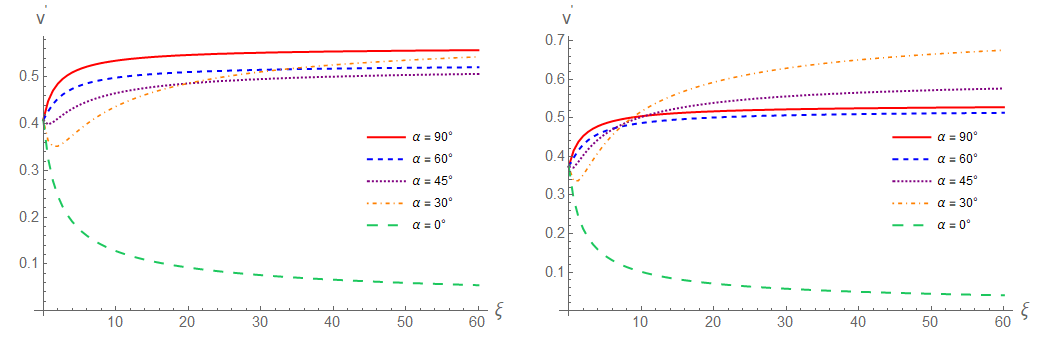}}
\caption{\small Plots of \( v'_\alpha (\xi) \) as a function of the anisotropy parameter \( \xi \) for $\alpha=\pi/2$ (solid),  $\alpha=2\pi/3$ (dashed), $\alpha=\pi/4$ (dotted), $\alpha=\pi/6$ (dashed-dotted)  and $\alpha=0$ (long dash) for  \(\sigma = 5\) (left) and \(\sigma = 10\) (right).}
\label{graph_v2}
\end{figure}

Let us now analyse the anisotropy dependence of the absolute value of downstream velocity $v'_\alpha (\xi) \equiv \vert {\bf v' (\xi \vert \alpha)} \vert $ at different incidence angles $\alpha$, $\alpha \in [0,\pi/2]$. The resulting curves  are shown in Fig.~\ref{graph_v2}  for the same values of $\sigma$ as in Fig.~\ref{graph_v1}.
The behaviour of $v'_\alpha (\xi)$ is characterised by two different patterns. The transition between them, similarly to the above-considered case of upstream velocity,  also takes place at $\alpha_{\rm crit} \sim \pi/4$:
\begin{itemize}
\item In the interval of incidence angles $\alpha \in (0,\pi/4)$ a  growth at large anisotropies  is preceded by the minimum at some $\xi^*(\alpha)$ such that $\xi^* (\alpha) \to 0$ at $\alpha \to \alpha_{\rm crit}$ and $\xi^* (\alpha) \to \infty$ at $\alpha \to 0$. A detailed illustration of this pattern is presented in Fig.~\ref{ris_v2_gap}. 
\item In the interval of incidence angles $\alpha \in (\pi/4,\pi/2)$, similarly to the behaviour of  $v_\alpha (\xi)$ in the same interval of angles, the function $v'_\alpha (\xi)$ is a concave growing one smoothly approaching the limiting curve for the transverse shock wave at at $\alpha=\pi/2$;
\begin{figure}[H]
\center{\includegraphics[width=1\linewidth]{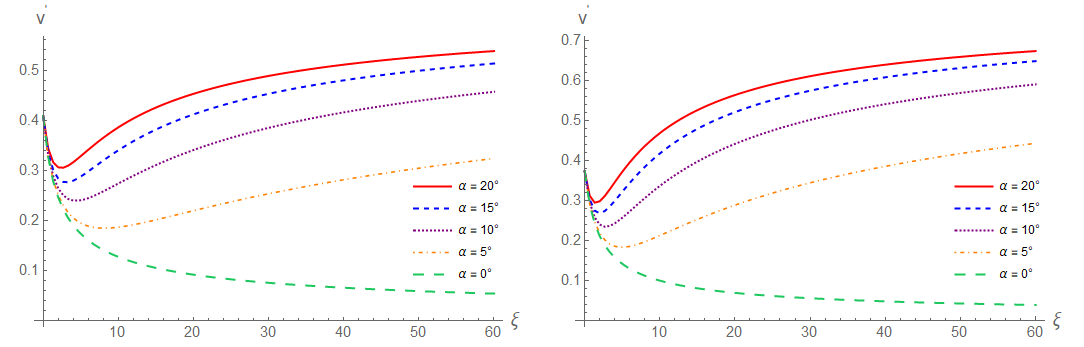}}
\caption{\small Plots of \( v'_\alpha (\xi) \) as a function of the anisotropy parameter \( \xi \) for $\alpha=20^{\circ}$ (solid), $\alpha=15^{\circ}$ (dashed), $\alpha=10^{\circ}$ (dotted), $\alpha=5^{\circ}$ (dash-dotted) and $\alpha=0$ (long dash) for  \(\sigma = 5\) (left) and \(\sigma = 10\) (right).}
\label{ris_v2_gap}
\end{figure}
Let us note that at small angles the form of $v'_\alpha (\xi)$ is extremely sensitive to the value of $\alpha$, see Fig.~\ref{ris_v2_end}, possibly indicating an unstable velocity transformation pattern of the "almost longitudinal" shock waves.
\begin{figure}[H]
\center{\includegraphics[width=1\linewidth]{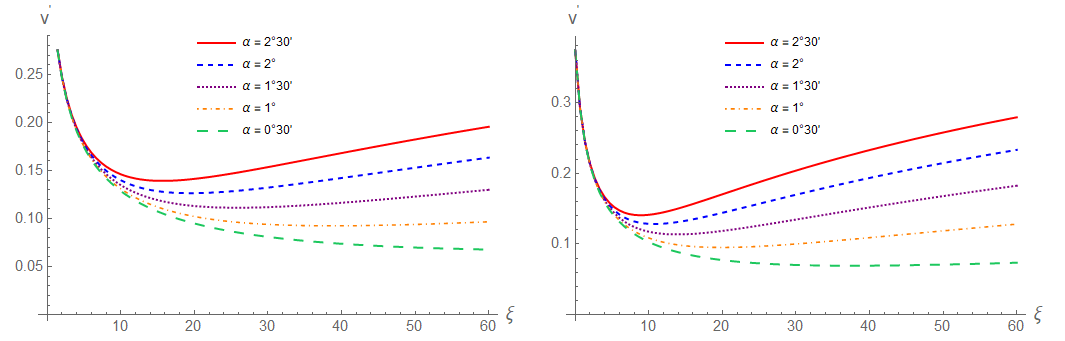}}
\caption{\small Plots of \( v'_\alpha (\xi) \) as a function of the anisotropy parameter \( \xi \) for small incidence angles  $\alpha=2^{\circ}30^{'}$ (solid), $\alpha=2^{\circ}$ (dashed), $\alpha=1^{\circ}30^{'}$ (dotted), $\alpha=1^{\circ}$ (dash-dotted) and $\alpha=0^{\circ}30^{'}$ (long dash) for  \(\sigma = 5\) (left) and \(\sigma = 10\) (right).}
\label{ris_v2_end}
\end{figure}
\end{itemize}

\subsubsection{Transformation $\vert {\bf v} \vert \to \vert {\bf v'} \vert $: angular dependence}

Let us now consider the angular dependence of the pattern of anisotropy dependence of the relative change of the absolute value of velocity  $\delta_{\alpha \alpha'} (\xi)$ defined in Eq.~\eqref{deltaalpha} induced by a superposition of the corresponding patterns for  $\vert {\bf v (\xi \vert \alpha)} \vert$  and $\vert {\bf v' (\xi \vert \alpha)} \vert$ studied in the previous paragraph \ref{vvp}.

\begin{figure}[H]
\center{\includegraphics[width=1\linewidth]{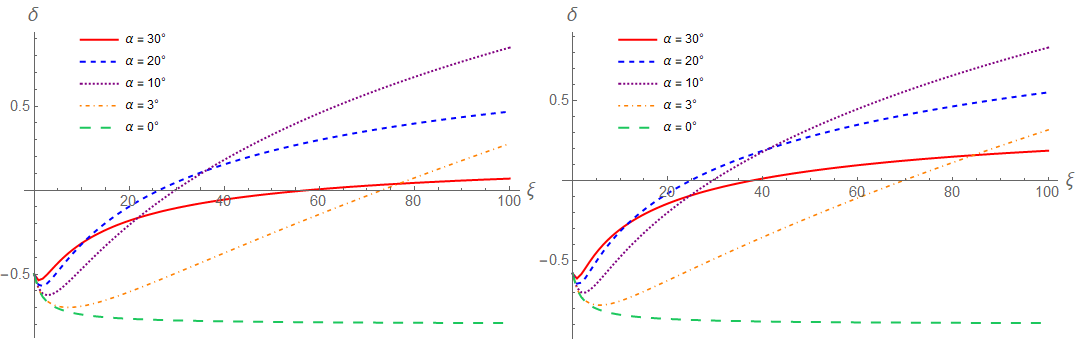}}
\caption{\small Plots of  $\delta_{\alpha \alpha'} (\xi) $  as a function of anisotropy parameter for $\alpha=30^{\circ}$ (solid), $\alpha=20^{\circ}$ (dashed), $\alpha=10^{\circ}$ (dotted), $\alpha=3^{\circ}$ (dash-dotted) and $\alpha=0$ (long dash)  for \(\sigma = 5\) (left) and \(\sigma = 10\) (right).}
\label{ris_delta}
\end{figure}

As seen at Figure \ref{ris_delta}, for each shock wave incidence angle $\alpha \in (0,\pi/2)$ at some critical anisotropy $\xi^* (\alpha)$ the relative velocity drop $\delta_{\alpha \alpha'} (\xi) $ changes its sign. This means that at sufficiently large anisotropies the rarefaction shock wave pattern with $\delta_{\alpha \alpha'} (\xi) <0$ turns into the compression shock wave one with $\delta_{\alpha \alpha'} (\xi) >0$ corresponding to acceleration of the flow by the shock wave so that we see a dramatic anisotropy-induced transition in the very nature of shock waves in anisotropic relativistic hydrodynamics. 

\begin{figure}[H]
\center{\includegraphics[width=1\linewidth]{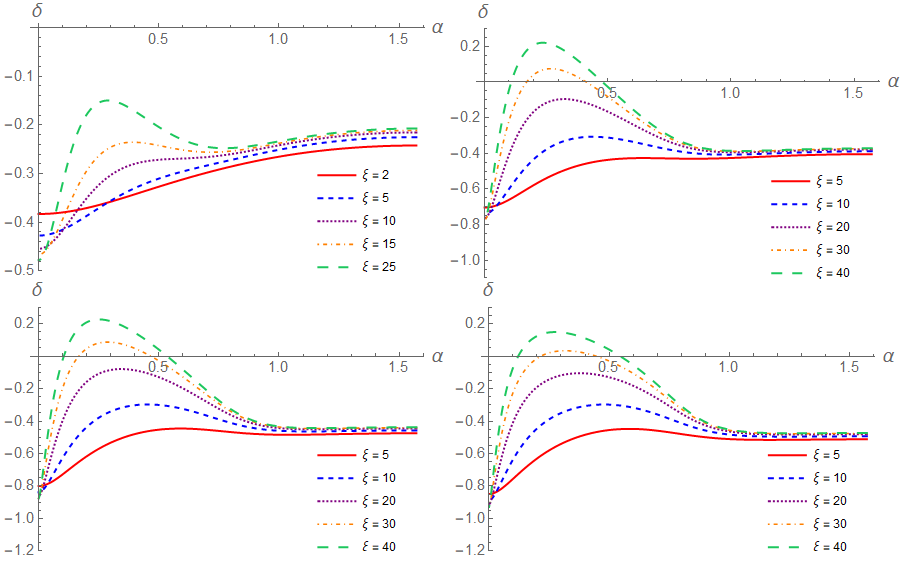}}
\caption{\small Plots of  $\delta_{\alpha \alpha'} (\xi) <0$  as a function of  the incidence angle \(\alpha\) for $\xi=2$ (solid), $\xi=5$ (dashed), $\xi=10$ (dotted), $\xi=15$ (dash-dotted) and $\xi=25$ (long dash) for \(\sigma = 2\) (top left), \(\sigma = 5\) (top right), \(\sigma = 10\) (bottom left) and \(\sigma = 20\) (bottom right).}
\label{ris_delta_alpha}
\end{figure}

\begin{figure}[H]
\center{\includegraphics[width=1\linewidth]{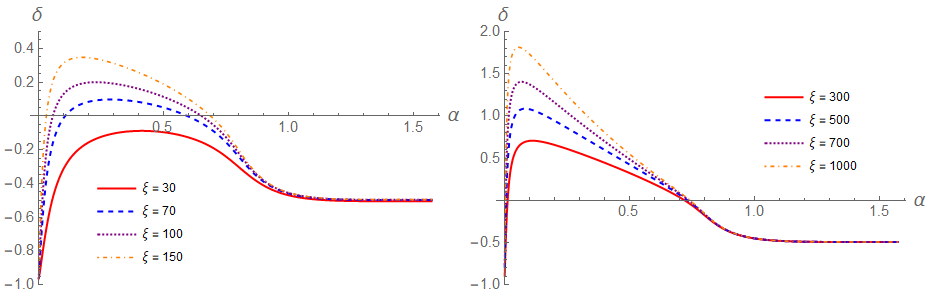}}
\caption{\small Plots of $\delta_{\alpha \alpha'} (\xi) $  as a function of the incidence angle \(\alpha\) for large values of the anisotropy parameter $\xi=30$ (solid), $\xi=70$ (dashed), $\xi=100$ (dotted) and $\xi=150$ (dash-dotted).}
\label{ris_delta_alpha_big}
\end{figure}

Let us now analyse the relative change of the absolute value of velocity  $\delta_{\alpha \alpha'} (\xi) $ as a function of the incidence angle $\alpha$ at fixed $\xi$ In Figs.~(\ref{ris_delta_alpha},\ref{ris_delta_alpha_big}) we plot this dependence for several relatively small (Fig.~(\ref{ris_delta_alpha})) and very large (Fig.~(\ref{ris_delta_alpha_big})) values of the anisotropy parameter and four different values of $\sigma$ in Fig.~(\ref{ris_delta_alpha}) and $\sigma=100$ in Fig.~(\ref{ris_delta_alpha_big}).

From Figs.~(\ref{ris_delta_alpha},\ref{ris_delta_alpha}) we see that the superposition of the angular  dependencies of $\vert {\bf v (\xi \vert \alpha)} \vert$  and $\vert {\bf v' (\xi \vert \alpha)} \vert$ leads to a hump-backed pattern for $\delta_{\alpha \alpha'} (\xi) $ with the hump moving from large to small angles and becoming more pronounced with increasing anisotropy. 

For large $\sigma$ and $\xi$ there appears an interval of angles in which $\delta_{\alpha \alpha'} (\xi) $ changes its sign and, therefore, a shock wave pattern changes from the rarefaction to the compression one. The width of this interval grows with $\sigma$.

\begin{figure}[H]
\center{\includegraphics[width=1\linewidth]{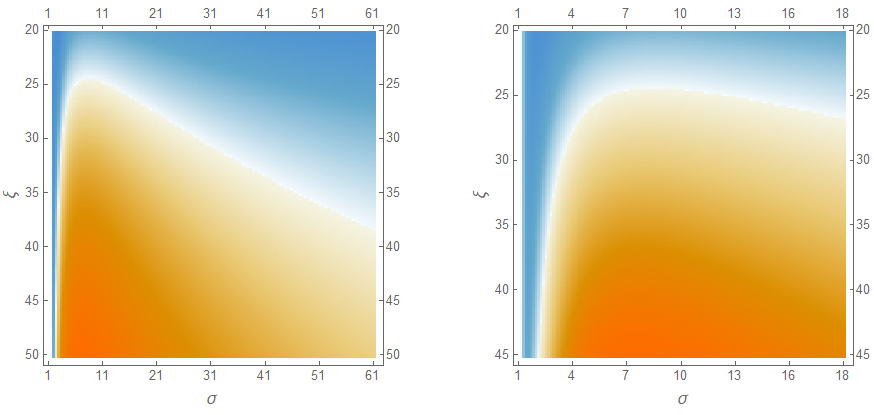}}
\caption{\small Plot of the location of the maximum of $\delta_{\alpha \alpha'} (\xi)$ . Warmer tones are positive values, but colder tones are negative values. Almost zero value corresponds to white color.}
\label{ris_min}
\end{figure}

In Fig.~\ref{ris_min} we plot a position of the hump in the $(\xi,\sigma)$ plane.
From Fig.~\ref{ris_min} we see that \(\delta (\alpha) \) touches zero  for two values of \(\sigma \). For small  \(\sigma \) and \(\xi\) the dependence is nonlinear.
For anisotropy parameters \(\xi\) below a certain value, as \(\sigma\) grows, \(\delta\) does twice undergo a transition between negative and positive ranges. Thus, for such values of \(\xi\), there are two possible types of shock waves with a feature typical for compression shock waves - a deceleration of the upstream flow. The first type of waves is characterised by  small values of \(\sigma\) while for the second type  \(\sigma\) takes large values that grow almost linearly with increasing \(\xi\).

\subsubsection{Transformation $(p_T,p_L) \to (p^{'}_T,p^{'}_L)$: angular dependence}

Of particular interest for describing effects of downstream and upstream flows related to shock wave formation for heavy ion collisions are the associated transverse and longitudinal momenta that contribute to transverse momentum and rapidity spectra.  For a shock wave incident at polar angle $\alpha$ the corresponding transverse and longitudinal momenta for the upstream flow read
\begin{equation}
p_T = p \sin \alpha, \, \, \, p_L = p \cos\alpha.
\end{equation}
Analogous formulae hold for the downstream flow. The resulting angular dependencies $p_T (\alpha \vert \xi)$, $p^{'}_T (\alpha^{'} \vert \xi)$, $p_L (\alpha \vert \xi)$ and 
 $p^{'}_L(\alpha^{'} \vert \xi)$ are shown, for several values of $\xi$, in Figs.~\ref{pT} and \ref{pL} correspondingly.
\begin{figure}[H]
\center{\includegraphics[width=1\linewidth]{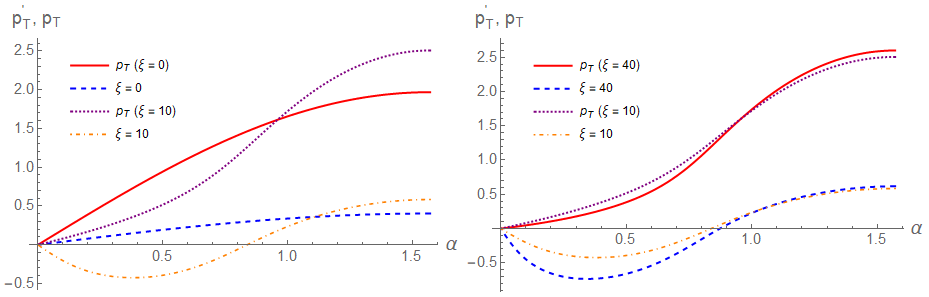}}
\caption{\small Left: plots of  $p_T (\alpha \vert \xi)$  for $\xi=0$ (solid),  $\xi=40$ (dotted) and $p^{'}_T (\alpha \vert \xi)$  for $\xi=0$ (dashed),  $\xi=10$ (dashed-dotted).   
Right: plots of  $p_T (\alpha \vert \xi)$  for $\xi=10$ (solid),  $\xi=40$ (dotted) and $p^{'}_T (\alpha \vert \xi)$  for $\xi=10$ (dashed),  $\xi=40$ (dashed-dotted).}
\label{pT}
\end{figure}
\begin{figure}[H]
\center{\includegraphics[width=1\linewidth]{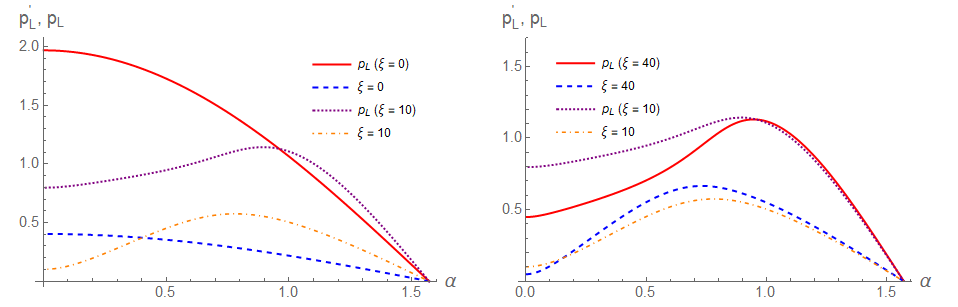}}
\caption{\small Left: plots of  $p_L (\alpha \vert \xi)$  for $\xi=0$ (solid),  $\xi=40$ (dotted) and $p^{'}_L (\alpha \vert \xi)$  for $\xi=0$ (dashed),  $\xi=10$ (dashed-dotted).   
Right: plots of  $p_L (\alpha \vert \xi)$  for $\xi=10$ (solid),  $\xi=40$ (dotted) and $p^{'}_L (\alpha \vert \xi)$  for $\xi=10$ (dashed),  $\xi=40$ (dashed-dotted).}
\label{pL}
\end{figure}

We see from Fig.~\ref{pT} that in comparison to isotropic production there takes place a reversion of $p^{'}_T$  at small incidence angles $\alpha < \alpha^*_T (\xi)$ and more $p^{'}_T$ is produced at large incidence angles $\alpha^*_T (\xi) < \alpha <\pi/2$ where $\alpha^*_T (\xi)$ is an anisotropy - dependent scale separating the regimes of transverse momentum reverssal at small $\alpha$ and its enrichment at large $\alpha$. 

As to the longitudinal momenta, in Fig.~\ref{pT} we observe, in comparison to the isotropic case, a pattern of depletion of longitudinal momentum at small angles $\alpha < \alpha^*_L (\xi)$ and its enhancement at large angles $\alpha^*_L (\xi) < \alpha <\pi/2$ where  $\alpha^*_L (\xi)$ is a regime-dividing scale \emph{ different} from its transverse counterpart $\alpha^*_L (\xi)$.

\section{Conclusions}

Let us summarise the main results obtained in the paper:
\begin{itemize}
\item General equations describing shock waves in relativistic anisotropic hydrodynamics were derived.
\item Solutions describing normal shock waves incident at an arbitrary angle with respect to collision axis as well as transverse and longitudinal shock waves were obtained and compared with the corresponding results for the isotropic case.
\item A new phenomenon of anisotropy - related angular deflection of the upstream flow was described.
\item Transformation of velocities and momenta by the shock wave front was analysed.
\end{itemize} 

In our view among the problems worth further studies the most interesting and pressing one is an analysis of entropy transformation by the shock wave front in relativistic anisotropic hydrodynamics. It is well known that anisotropy gives rise to a new source of entropy production in anisotropic hydrodynamics so it is very interesting to see how the standard pattern of entropy production by shock waves in isotropic hydrodynamics changes in the anisotropic case. We plan to address this problem in the near future.

\begin{center}
{\bf Acknowledgements}
\end{center}

The work was supported by the RFBR Grant 18-02-40069

The authors have no conflicts of interest to declare that are relevant to the content of this article.

\printbibliography

%\appendix*
\section*{Appendix A. Subluminality condition for shock waves in relativistic anisotropic hydrodynamics}

In the anisotropic case the necessary subluminality condition $N^\mu N_\mu<0$ for the four-vector $N_\mu$ orthogonal to the discontinuity surface can be studied by writing the corresponding equations generalising equations (\ref{n1},\ref{n2}) for the isotropic case. 

From the matching conditions  (\ref{gap}) and the expression (\ref{T_true}) for the energy momentum tensor one gets the following system of equations:
\begin{align}
(\varepsilon + P_\perp^{'}) x &= (\varepsilon^{'} + P_\perp^{'}) A x^{'} - (P_\perp^{'} - P^{'}_\parallel) C y^{'},
\label{a1_appendix}
\\
(\varepsilon^{'} + P_\perp) x^{'} &= (\varepsilon + P_\perp) A x - (P_\perp - P_\parallel) B y,
\label{a2_appendix}
\\
(P_\perp^{'} - P_\parallel) y &= (\varepsilon^{'} + P_\perp^{'}) B x^{'} - (P_\perp^{'} - P^{'}_\parallel) D y^{'},
\label{a3_appendix}
\\
(P_\perp - P_\parallel^{'}) y^{'} &= (\varepsilon + P_\perp) C x - (P_\perp - P_\parallel) D y,
\label{a4_appendix}
\end{align}
where
\begin{align}
& x = U_\mu N^\mu, \ \ \ \ \ x^{'} = U^{'}_\mu N^\mu, \ \ \ \ \ A = U^{'}_\mu U^\mu, \ \ \ \ \ B = U^{'}_\mu Z^\mu,\\
& y = Z_\mu N^\mu, \ \ \ \ \ y^{'} = Z^{'}_\mu N^\mu, \ \ \ \ \ C = Z^{'}_\mu U^\mu, \ \ \ \ \ D = Z^{'}_\mu Z^\mu.
\end{align}

In ultra-relativistic case we have
\begin{align}
(2P_\perp + P_\parallel + P_\perp^{'}) x &= (3P_\perp^{'} + P_\parallel^{'}) A x^{'} - (P_\perp^{'} - P^{'}_\parallel) C y^{'},
\label{a1_n}
\\
(2P_\perp^{'} + P_\parallel^{'} + P_\perp) x^{'} &= (3P_\perp + P_\parallel) A x - (P_\perp - P_\parallel) B y,
\label{a2_n}
\\
(P_\perp^{'} - P_\parallel) y &= (3P_\perp^{'} + P_\parallel^{'}) B x^{'} - (P_\perp^{'} - P^{'}_\parallel) D y^{'},
\label{a3_n}
\\
(P_\perp - P_\parallel^{'}) y^{'} &= (3P_\perp + P_\parallel) C x - (P_\perp - P_\parallel) D y,
\label{a4_n}
\end{align}

Our purpose is to evaluate the sign of \(N_\mu N^\mu\) for check subluminality condition, which is $N^\mu N_\mu<0$. From  (\ref{gap_expand_appendix}) one can find
\begin{align}
N_\mu N^\mu = \frac{1}{P_\perp - P^{'}_\perp} \Big[ (3P_\perp + P_\parallel) x^2 - (3P_\perp^{'} + P_\parallel^{'}) x^{'2} - (P_\perp - P_\parallel) y^2 + (P_\perp^{'} - P^{'}_\parallel) y^{'2} \Big].
\label{N_norm_aniso}
\end{align}

The system of equations (\ref{a1_n} - \ref{a4_n}) has solutions if the condition for the determinant is satisfied.
\begin{equation}
\Det \Lambda = 
\begin{vmatrix}
0 & - (3P_\perp^{'} + P_\parallel^{'}) B & (P_\perp^{'} - P_\parallel)  & (P_\perp^{'} - P^{'}_\parallel) D \\ 
- (3P_\perp + P_\parallel) C & 0 & (P_\perp - P_\parallel) D & (P_\perp - P_\parallel^{'})  \\ 
(2P_\perp + P_\parallel + P_\perp^{'}) & -(3P_\perp^{'} + P_\parallel^{'}) A & 0 & (P_\perp^{'} - P^{'}_\parallel) C  \\ 
-(3P_\perp + P_\parallel) A & (2P_\perp^{'} + P_\parallel^{'} + P_\perp) & (P_\perp - P_\parallel) B & 0 
\end{vmatrix}
= 0.
\label{det}
\end{equation}

It is worth saying that in the borderline cases, then the flow moves along the axes \(Ox \) and \(Oz \), the equation \(Det \Lambda = 0 \) gives the correct solutions for \(\Delta \), which agree with solutions for the velocities in anisotropic case.

Then, solving the system of equations, we can express \( x^{'}, y, y^{'}\) through \(x\), that give us
\begin{align}
N_\mu N^\mu &= \frac{1}{P_\perp - P^{'}_\perp} \Phi (P_\perp, P_\parallel, P^{'}_\perp, P^{'}_\parallel, A, B, C, D) x^2
\end{align}

In a case of an arbitrary polar angle due to the symmetry in \(Oxy\)-plane we can fix \(x\)-direction and consider movement only in \(Oxz\)-plane:
\begin{align}
U_\mu &= (u_0 \cosh \vartheta, u_x, 0, u_0 \sinh \vartheta),\\
Z_\mu &= (\sinh \vartheta, 0,0, \cosh \vartheta).
\end{align}

Denote
\begin{align}
u_x &= \sinh \gamma, \ \ u_0 = \cosh \gamma, \\
u^{'}_x &= \sinh \gamma^{'}, \ \ u^{'}_0 = \cosh \gamma^{'}.
\end{align}
Then we will hasve for \(A, \ B, \ C and D\) the following formulas:
\begin{align}
A &= U^{'}_\mu U^\mu =  \cosh \gamma \cosh \gamma^{'} \cosh (\vartheta - \vartheta^{'}) - \sinh \gamma \sinh \gamma^{'}
\\
B &= U^{'}_\mu Z^\mu = \cosh \gamma^{'} \sinh (\vartheta - \vartheta^{'}),
\\
C &= Z^{'}_\mu U^\mu = - \cosh \gamma \sinh (\vartheta - \vartheta^{'})
\\
D &= Z^{'}_\mu Z^\mu = - \cosh (\vartheta - \vartheta^{'}).
\end{align}

Thus, the system of equations will explicitly depend only on the difference \(\vartheta - \vartheta ^ {'}\), and not on the values themselves. Let's denote \(\Delta = \vartheta - \vartheta^{'} \).

Due to the anisotropic hydrodynamic \(P_\perp, \ P_\parallel\) are divided into anisotropic and anisotropic parts according to the formulas (\ref{pT}, \ref{pL}). Also denote \(\sigma = P^{'}_{iso} / P_{iso}\). From the equation \(\Det \Lambda = 0 \) we can get the value \(\Delta \). Thus, we have 4 unknowns \(\sigma, \xi, \gamma, \gamma^{'} \), of which \(\sigma, \xi \) are the parameters of the system. Also, solving the equation \(\Det \Lambda = 0 \) allow us to obtain a consistent system of equations (\ref{a1_n} - \ref{a4_n}), so it is possible to choose one among the values \(x, x^{'}, y, y^{'}\), and express others through it. Let, for example, this be the value \(x\), then, from the expression (\ref{N_norm_aniso}), we can write

\begin{align}
N_\mu N^\mu &= \frac{1}{R_\perp(\xi) (1 - \sigma)} \Phi (\sigma, \xi, \gamma, \gamma^{'}, \Delta) x^2.
\label{N_norm_aniso3}
\end{align}

One can construct the following function
\begin{align}
S (\sigma, \xi, \gamma, \gamma^{'}) = \Sign \Bigg( \frac{1}{R_\perp(\xi) (1 - \sigma)} \Phi \Big(\sigma, \xi, \gamma, \gamma^{'}, \Delta (\sigma, \xi, \gamma, \gamma^{'}) \Big) \Bigg),
\end{align}
that determines the sign of the norm of the vector \(N^\mu\).

Denote \(T = \tanh \gamma, T^{'} = \tanh \gamma^{'}\) and then he have
\begin{align}
S (\sigma, \xi, T, T^{'}) = \Sign \Bigg( \frac{1}{R_\perp(\xi) (1 - \sigma)} \Phi \Big(\sigma, \xi, T, T^{'}, \Delta (\sigma, \xi, T, T^{'}) \Big) \Bigg),
\end{align}

It should be taken into account that the equation \(\Det \Lambda = 0 \) may not have solutions, then it is convenient to construct the following function:
\begin{equation}
\Omega (\sigma, \xi, T, T^{'}) = 
 \begin{cases}
   -1 &\text{$\Det \Lambda = 0$  has no solutions}\\
   0 &\text{$S (\sigma, \xi, \gamma, \gamma^{'}) = -1$}\\
   1 &\text{$S (\sigma, \xi, \gamma, \gamma^{'}) = 1$}\\
 \end{cases}
\end{equation}

\begin{figure}[H]
\center{\includegraphics[width=0.99\linewidth]{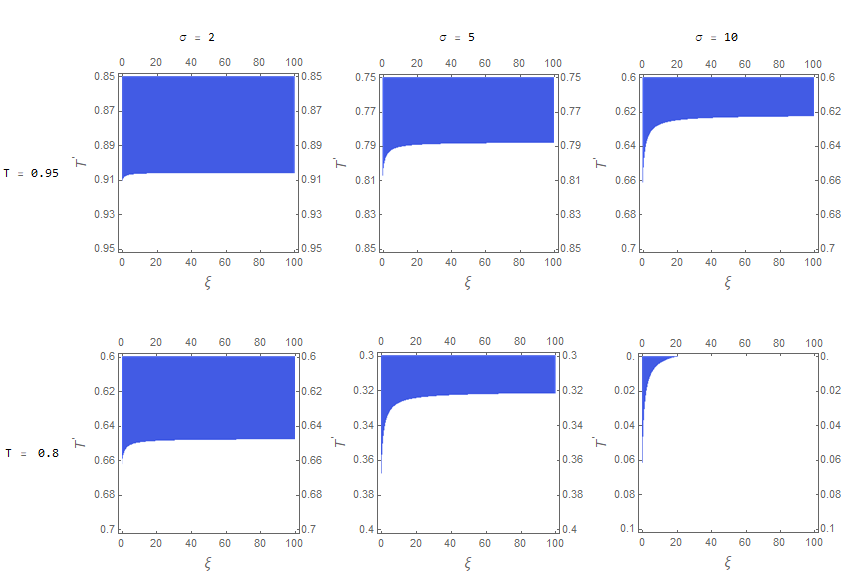}}
\caption{\small Plot of \(\Omega (\sigma, \xi, T, T^{'})\) as a function of \(\xi,\ T^{'}\). Blue color denotes \(\Omega = -1\), white color - \(\Omega = 0\).}
\label{sub}
\end{figure}

As one can see on Figure \ref{sub}, it turned out that there is no case of \(\Omega = 1\) anywhere. It can also be seen that the graphs show as \(\xi \) increases, the region of possible solutions to the equation \(\Det \Lambda = 0 \) increases. This fact allows us to assume that in an anisotropic medium, a shock wave can be formed more often than in an isotropic case.

	\section*{Appendix B}
	
	Let us first consider a description of a shock wave propagating perpendicular to the beam axis. Due to the symmetry in \(Oxy\)-plane, it sufficient to consider its propagation along the $x$ axis and, correspondingly, choose the following basis:
\begin{align}
U_\mu &= (u_0, u_x, 0, 0), \, \, \, Z_\mu = (0, 0, 0, 1),
\label{basis1}
\\
U^{'}_\mu &= (u^{'}_0, u^{'}_x, 0, 0), \, \, \, Z^{'}_\mu =  Z_\mu = (0, 0, 0, 1).
\label{basis2}
\end{align}

The matching conditions (\ref{gap}) lead to the following system of equations: 
\begin{align}
& (\varepsilon + P_\perp)U_0 x - P_\perp N_0 - (\varepsilon^{'} + P^{'}_\perp)U^{'}_0 x^{'} + P^{'}_\perp N_0 = 0, \\
& (\varepsilon + P_\perp)U_1 x - P_\perp N_1 -(\varepsilon^{'} + P^{'}_\perp)U^{'}_1 x^{'} + P^{'}_\perp N_1 = 0, \\
&  - P_\perp N_2 + P^{'}_\perp N_2 = 0, \\
&  - P_\parallel N_3 + P^{'}_\parallel N_3 = 0.
\label{eqs}
\end{align}

The third and fourth equations lead us to the solutions \(N_2 = 0\) and \(N_3 = 0\). We also take into account the expression for the ultrarelativistic case and obtain
\begin{align}
& (3P_\perp + P_\parallel)u_0 \, (u_0  N_0 - u_x N_1) - P_\perp N_0- (3P^{'}_\perp + P^{'}_\parallel)u^{'}_0\,  (u^{'}_0 N_0 - u^{'}_x N_1)  + P^{'}_\perp N_0 = 0,
\label{eqs2_1}
\\
& (3P_\perp + P_\parallel)u_x  (u_0 N_0 - u_x N_1)  - P_\perp N_1  -(3P^{'}_\perp + P^{'}_\parallel)u^{'}_x (u^{'}_0 N_0 - u^{'}_x N_1) + P^{'}_\perp N_1 = 0.
\label{eqs2_2}
\end{align}

For the existence of solutions to the remaining two equations on \(N_0, N_1 \), the determinant of the coefficients of the equation should be equal to zero. With introducing the following definitions
\begin{align}
A_0 &= (3P_\perp + P_\parallel)u^2_0 - (3P^{'}_\perp + P^{'}_\parallel)u^{'2}_0 - (P_\perp - P^{'}_\perp), 
\label{koeff_start}
\\
B_0 &= - (3P_\perp + P_\parallel)u_0 u_x + (3P^{'}_\perp + P^{'}_\parallel)u^{'}_0 u_x^{'}, \\
A_1 &= - B_0 = (3P_\perp + P_\parallel)u_0 u_x - (3P^{'}_\perp + P^{'}_\parallel)u^{'}_0 u_x^{'} , \\
B_1 &= - (3P_\perp + P_\parallel)u^2_x + (3P^{'}_\perp + P^{'}_\parallel)u^{'2}_x - (P_\perp - P^{'}_\perp), \\
\label{koeff_end}
\end{align}
determinant will take the form
\begin{equation}
\begin{vmatrix}
A_0 & B_0  \\ 
A_1 & B_1  
\end{vmatrix}
= 0.
\label{det}
\end{equation}

The one-dimensional formulation of the problem allows us to write expressions for the components of the 4-velocity vector in terms of hyperbolic functions
\begin{align}
u_x &= \sinh \gamma, \ \ u_0 = \cosh \gamma, \\
u^{'}_x &= \sinh \gamma^{'}, \ \ u^{'}_0 = \cosh \gamma^{'}.
\end{align}
Substituting (\ref{det}) and (\ref{koeff_start} - \ref{koeff_end}) into (\ref{det}) we obtain
\begin{align}
&(3P_\perp + P_\parallel)(3P^{'}_\perp + P^{'}_\parallel) \cosh (2\gamma - 2\gamma^{'}) = P_\perp (P_\parallel^{'} + 4P_\perp ) + \nonumber \\ &+ P_\perp^{'} (2P_\parallel^{'} + P_\perp) + 4P_\perp^{'2} + P_\parallel (P_\parallel^{'} + 2P_\perp + P_\perp^{'} ),
\label{det_expand}
\end{align}
or, more conveniently, as
\begin{equation}
\sinh^2 (\gamma - \gamma^{'}) = \frac{(2P_\perp - P_\parallel - 2P^{'}_\perp + P^{'}_\parallel)(P_\parallel - P^{'}_\parallel)}{(3P_\perp + P_\parallel)(3P^{'}_\perp + P^{'}_\parallel)}.
\end{equation}

In the limit $\xi \to 0$ the forluma above is equal to the solution in the isotropic case \cite{Mitchell}. Space-like nature of the normal vector \(N^\mu\), i.e \(N^\mu N_\mu = -1 \), lead to a closed system of equations for \(N_0, N_1\), solutions of whose are
\begin{align}
N_0 &= \frac{1}{2} \frac{(3P_\perp + P_\parallel) \cosh (2\gamma - \gamma^{'}) - (P_\parallel^{'} + P_\perp + 2P_\perp^{'})\cosh \gamma^{'}}{\sqrt{(P_\parallel + 2P_\perp + P_\perp^{'})^2 \sinh^2 (\gamma - \gamma^{'}) - (P_\parallel - P^{'}_\parallel)^2 \cosh^2 (\gamma - \gamma^{'})}},
\label{n0_general}
\\
N_1 &= \frac{1}{2} \frac{(3P_\perp + P_\parallel) \sinh (2\gamma - \gamma^{'}) - (P_\parallel^{'} + P_\perp + 2P_\perp^{'})\sinh \gamma^{'}}{\sqrt{(P_\parallel + 2P_\perp + P_\perp^{'})^2 \sinh^2 (\gamma - \gamma^{'}) - (P_\parallel - P^{'}_\parallel)^2 \cosh^2 (\gamma - \gamma^{'})}},
\label{n1_general}
\end{align}

Similarly to the example from relativistic hydrodynamics, consider the case when the normal vector is directed along the \(Ox\)-axis \(N^\mu = (0,1,0,0 )\), then from the equations (\ref{eqs2_1} - \ref{eqs2_2}) we find expressions for the velocities \(u_x, \ u_x^{'}\):
\begin{align}
&\sinh \gamma = u_x = \sqrt{\frac{(P_\perp - P^{'}_\perp) (P_\parallel^{'} + P_\perp + 2P_\perp^{'})}{(2P_\perp - P_\parallel - 2P^{'}_\perp + P^{'}_\parallel)(P_\parallel^{'} + 2P_\perp + P_\perp^{'})}},
\label{u1}
\\
&\sinh \gamma^{'} = u_x^{'} = \sqrt{\frac{(P_\perp - P^{'}_\perp)(P_\parallel^{'} + 2P_\perp + P_\perp^{'})}{(2P_\perp - P_\parallel - 2P^{'}_\perp + P^{'}_\parallel)(P_\parallel^{'} + P_\perp + 2P_\perp^{'})}},
\label{u2}
\end{align}

\end{document}